\documentclass[prl,twocolumn,amssymb]{revtex4}
\usepackage{graphicx}

\begin{document}

\title{Spin-Orbit Coupled Spinor Bose-Einstein Condensates}
\author{Chunji Wang, Chao Gao, Chao-Ming Jian and Hui Zhai}
\email{hzhai@mail.tsinghua.edu.cn}
\affiliation{Institute for Advanced Study, Tsinghua University, Beijing, 100084, China}
\date{\today}

\begin{abstract}
An effective spin-orbit coupling can be generated in cold atom system by engineering atom-light interactions. In this letter we study spin-$1/2$ and spin-$1$ Bose-Einstein condensates with Rashba spin-orbit coupling, and find that the condensate wave function will develop non-trivial structures. From numerical simulation we have identified two different phases. In one phase the ground state is a single plane wave, and often we find the system splits into domains and an array of vortices plays the role as domain wall. In this phase, time-reversal symmetry is broken. In the other phase the condensate wave function is a standing wave and it forms spin stripe. The transition between them is driven by interactions between bosons. We also provide an analytical understanding of these results and determines the transition point between the two phases.
\end{abstract}
\maketitle

Interaction between matters field and gauge field is of central importance in quantum physics. Although atoms are neutral and do not possess gauge coupling to real electromagnetic field, it has been proposed that a synthetic gauge field originated from Berry phase effect can be coupled to atoms by engineering the interactions between atoms and a spatially varying laser field \cite{proposals}. Very recently, this scheme has been successfully implemented in a Rb$^{87}$ Bose-Einstein condensate (BEC) \cite{Ian}. A straightforward generalization of this scheme can also create a non-abelian gauge field to cold atoms \cite{non-abelian,spin-orbital,dalibard}. This opens up a new avenue in cold atom physics, that is to study how a coherent matter wave responses to external gauge fields, in particular, the non-abelian gauge fields.

There are already various proposals to achieve a non-abelian gauge field. For instance, one can start with a $N$-pod system where $N$ atomic internal states are coupled to one common state by $N$ different laser fields. It results in $N-1$ dark states which play a role as internal pseudo-spin degree of freedom \cite{non-abelian,spin-orbital,dalibard,spielman}. Tripod ($N=3$) and tetrapod ($N=4)$ setup correspond to spin-$1/2$ and spin-$1$ case, respectively. Among all possibly achieved configurations of non-abelian gauge fields ${\bf A}$, one of the most simplest case with $A_x=\sigma_x$ and $A_y=\sigma_y$ is equivalent to a Rashba type spin-orbit (SO) coupling \cite{spin-orbital,dalibard}. Very recently, BEC with SO coupling has been first realized by NIST group using a similar scheme described in Ref. \cite{Ian} \cite{spielman}.

During last a few years, it has been extensively studied that the SO coupling in an electronic system can lead to a novel state of matter of topological insulator which has many exotic physical properties \cite{TI}. Nevertheless, so far only few papers have studied the SO effect in a boson condensate \cite{BEC,Wu}. This letter is devoted to study the properties of spin-$1/2$ and spin-$1$ condensate in presence of a pure Rashba SO coupling. The model under consideration is $\hat{H}=\hat{H}_0+\hat{H}_{\text{int}}$,
\begin{equation}
\hat{H}_0=\int d^2{\bf r}\Psi^\dag\frac{1}{2m}\left({\bf k}^2+2\kappa {\bf k}\cdot {\vec \sigma}\right)\Psi,\label{single}
\end{equation}
where $\Psi=(\Psi_{\uparrow},\Psi_{\downarrow})$ for spin-$1/2$ case and $\Psi=(\Psi_1,\Psi_0,\Psi_{-1})$ for spin-$1$ case. Here we consider a quasi-two-dimensional situation and in-plane SO coupling where ${\bf k}=\{k_x,k_y\}$ and $\vec{\sigma}=\{\sigma_x,\sigma_y\}$, where $\sigma_{x,y}$ are spin-$1/2$ or spin-$1$ representation of Pauli matrices. For the interaction term, we note that different pseudo-spin components are in fact superposition of atomic hyperfine states, and therefore the interactions between them could have a quite complicated form. However, as an initial effort to understand this rich system, in this work we try to simplify the situation by considering a simplify interaction form borrowed from conventional spinor BEC, namely, for spin-$1/2$
\begin{equation}
H_{\text{int}}=\int d^2{\bf r}\left(g_1\hat{n}^2_\uparrow+g_2\hat{n}^2_\downarrow+2g_{12}\hat{n}_\uparrow \hat{n}_\downarrow \right).\label{spinhalf}
\end{equation}
we shall also focus on the case with $g_1=g_2>0$ and then the interaction can also be rewritten as
\begin{equation}
H_{\text{int}}=\int d^2{\bf r} \left(\frac{c_0}{2}\hat{n}^2+\frac{c_2}{2}\hat{S}_z^2\right)
\end{equation}
where $\hat{n}=\hat{n}_{\uparrow}+\hat{n}_{\downarrow}$, $\hat{S}_z=\hat{n}_{\uparrow}-\hat{n}_{\downarrow}$, and $c_0=g_1+g_{12}$ and $c_2=g_1-g_{12}$. The $c_0$ term is $SU(2)$ spin rotational invariant while the $c_2$ term breaks spin rotation symmetry. For spin-$1$ case, we shall consider the standard interaction form \cite{Ho}
\begin{equation}
H_{\text{int}}=\int d^2{\bf r} \left(\frac{c_0}{2}\hat{n}^2+\frac{c_2}{2}\hat{{\bf S}}^2\right)
\end{equation}
where $\hat{n}=\hat{n}_{1}+\hat{n}_0+\hat{n}_{-1}$ and $\hat{{\bf S}}=\Psi^\dag_\alpha{\vec \sigma}_{\alpha\beta}\Psi_\beta$. In both cases, $c_0>0$ and $\gamma=c_2/c_0$ can be either positive or negative. 

\begin{figure}[tbp]
\includegraphics[height=1.7in, width=2.4in]
{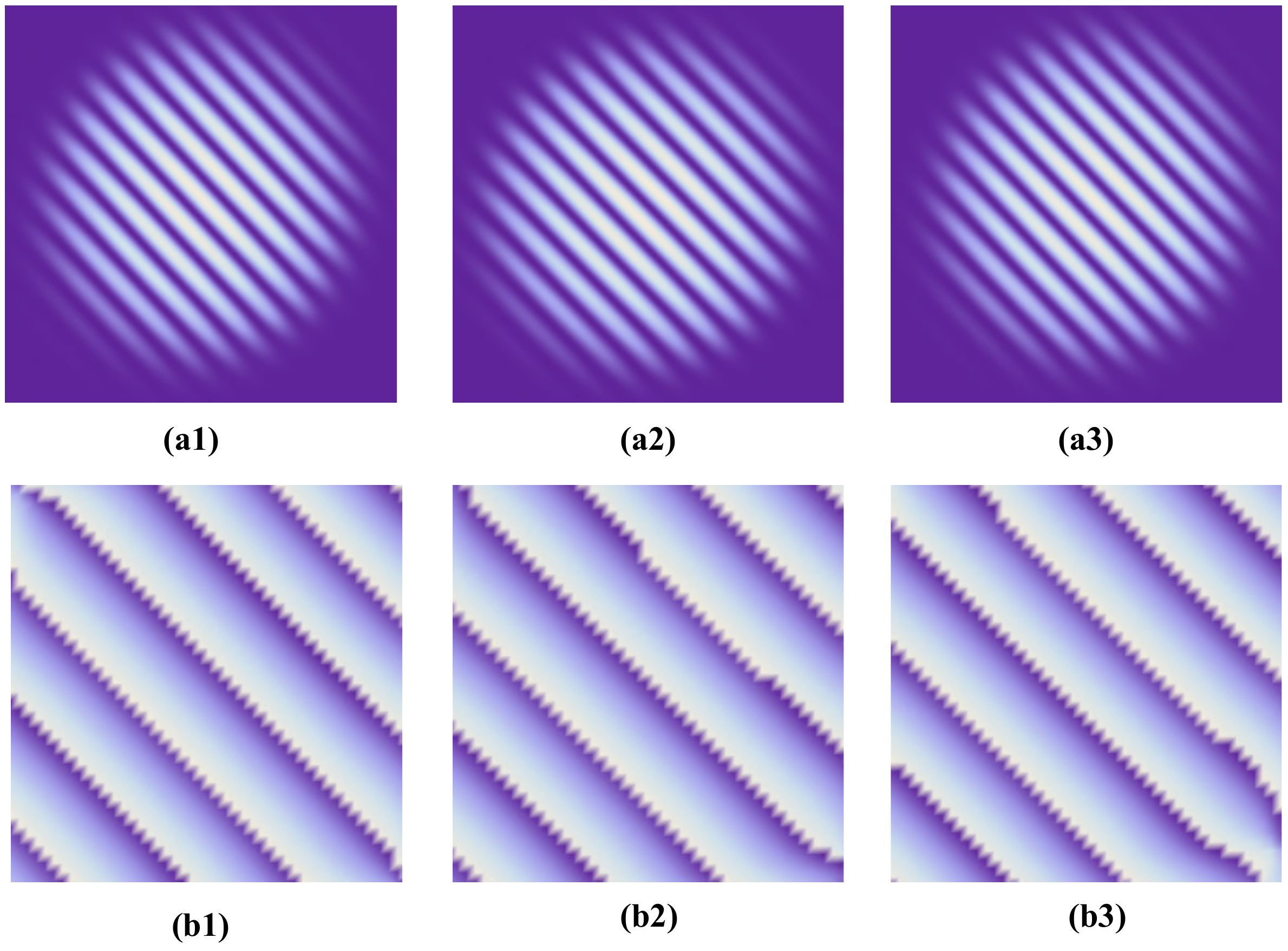} \caption{(Color online) Numerical results for spin-$1$ case. a1-a3 are density of $1$, $0$, and $-1$ component in ``SW " regime; and b1-b3 are phase of  and b3 are $1$, $0$, and $-1$ component in ``PW" regime. $\gamma=c_2/c_0=0.2$ for a1-a3, and $\gamma=-0.2$ for b1-b3.  \label{spin1}}
\end{figure}

The main results of this letter include: ({\bf I}) A boson condensation in ${\bf k}=0$ single particle state is always unstable in presence of SO coupling. ({\bf II}) The ground state has two possible phases: one is named as ``plane wave phase" (PW) where the ground state is found to be a single plane wave; the other is named as ``standing wave phase" (SW) where the spatial wave function of each spin component forms an oscillating standing wave, as presented in Fig. \ref{spinhalf} and \ref{spin1}. ({\bf III}) The transition between PW phase and SW phase depends on the interactions between bosons. For spin-$1/2$ case, it is PW phase if $\gamma>0$ and SW phase otherwise; while for spin-$1$ case, it is PW phase if $\gamma<0$ and SW phase otherwise. ({\bf IV}) We often find long-lived metastable state in which the system splits into two domains. In the parameter regime of PW phase, it is locally a plane wave state with opposite wave vector in each domain, and an equally-spaced array of vortices plays the role as domain wall. We note, unlike BEC in a synthetic magnetic field \cite{Ian}, the Hamiltonian for non-abelian gauge field considered here preserves both time-reversal and translation symmetry, however, the ``PW" phase spontaneously breaks time-reversal symmetry, which is unconventional in a bosonic system. Moreover, both two phases spontaneously break space-spin rotation symmetry. This behavior is fundamentally different from the effect of SO coupling in electronic systems. We have also verified that other terms not considered here, such as linear and quadratic Zeeman field, the difference between $g_1$ and $g_2$, will only shift the phase boundary between these two phases, and will not affect the essence of the two phases, as long as their strength is relatively not too strong.

{\it Instability of a condensation on ${\bf k}=0$ state:} Without SO coupling, the ground state of bosons is a condensation in ${\bf k}=0$ state. Hence, it arises the question whether the SO coupling ${\bf k}\cdot{\vec \sigma}$ will have any significant effect on a BEC, in particular, if the SO coupling is weak. Here let us take spin-$1/2$ case as an example, and first expand the boson field operator $\Psi^\dag_{\sigma}({\bf r})=\sum_{\bf k} e^{i{\bf k}{\bf r}}b^\dag_{{\bf k}\sigma}/\sqrt{V}$ ($V$ is the volumn). If we still assume a boson condensation in ${\bf k}=0$ state, i.e. $\langle b_{\sigma,{\bf k=0}}\rangle=\phi_{\sigma}$, the Bogoliubov Hamiltonian can be written as $\hat{H}_{\text{Bg}}=\sum_{{\bf k}}B^\dag_{{\bf k}}H_{{\bf k}}B_{{\bf k}}$
where
$B^\dag_{{\bf k}}=(\hat{b}^\dag_{\uparrow {\bf k}},  \hat{b}_{\uparrow {\bf -k}}, \hat{b}^\dag_{\downarrow {\bf k}}, \hat{b}_{\downarrow {\bf -k}})$ and
\begin{widetext}
\begin{eqnarray}
H_{{\bf k}}=\left(\begin{array}{cccc} \epsilon_{{\bf k}\uparrow} & 2g_1\phi^2_{\uparrow}& \frac{\hbar\kappa}{m}(k_x-ik_y)+2g_{12}\phi^*_{\uparrow}\phi_\downarrow & 2g_{12}\phi_{\uparrow}\phi_{\downarrow} \\2g_1\phi^{* 2}_{\uparrow} & \epsilon_{{\bf k}\uparrow}  & 2g_{12} \phi^*_{\uparrow}\phi^*_{\downarrow} \ & \frac{\hbar\kappa}{m}(-k_x-ik_y)+2g_{12}\phi^*_{\uparrow}\phi_\downarrow \\\frac{\hbar\kappa}{m}(k_x+i k_y)+2g_{12}\phi_{\uparrow}\phi^*_\downarrow & 2g_{12}\phi_{\uparrow}\phi_{\downarrow}  & \epsilon_{{\bf k}\downarrow} & 2g_2\phi^2_{\downarrow} \\2g_{12} \phi^*_{\uparrow}\phi^*_{\downarrow} & \frac{\hbar\kappa}{m}(-k_x+i k_y)+2g_{12}\phi_{\uparrow}\phi^*_\downarrow & 2g_2\phi^{* 2}_{\downarrow}  & \epsilon_{{\bf k}\downarrow} \end{array}\right)\nonumber
\end{eqnarray}
\end{widetext}
where $\epsilon_{{\bf k}\sigma}={\bf k}^2/(2m)+2g_1|\phi_{\sigma}|^2$. We then introduce a generalized Bogoliubov transformation $B_{{\bf k}}=\Omega_{{\bf k}}\tilde{B}_{{\bf k}}$, where $\tilde{B}^\dag_{{\bf k}}=(\hat{\beta}^\dag_{\uparrow{\bf k}}, \hat{\beta}_{\uparrow {\bf -k}}, \hat{\beta}^\dag_{\downarrow {\bf k}}, \hat{\beta}_{\downarrow{\bf -k}})$
where $\Omega_{{\bf k}}$ is a $4\times 4$ matrix. To satisfy the commutation relation $[\beta_{\sigma{\bf k}},\beta^\dag_{\sigma^\prime {\bf k^\prime}}]=\delta_{\sigma\sigma^\prime}\delta_{{\bf k}{\bf k^\prime}}$ and $[\beta_{\sigma{\bf k}},\beta_{\sigma^\prime{\bf k^\prime}}]=0$, $\Omega_{{\bf k}}$ has to satisfy the relation $\Omega^\dag_{{\bf k}} A\Omega_{{\bf k}}=A,\label{OmegaA}$ where $A$ is a $4\times 4$ matrix
$\left(\begin{array}{cc}\sigma_z & 0  \\0 & \sigma_z \end{array}\right)$.

\begin{figure}[bp]
\includegraphics[height=1.2in, width=3.0in]{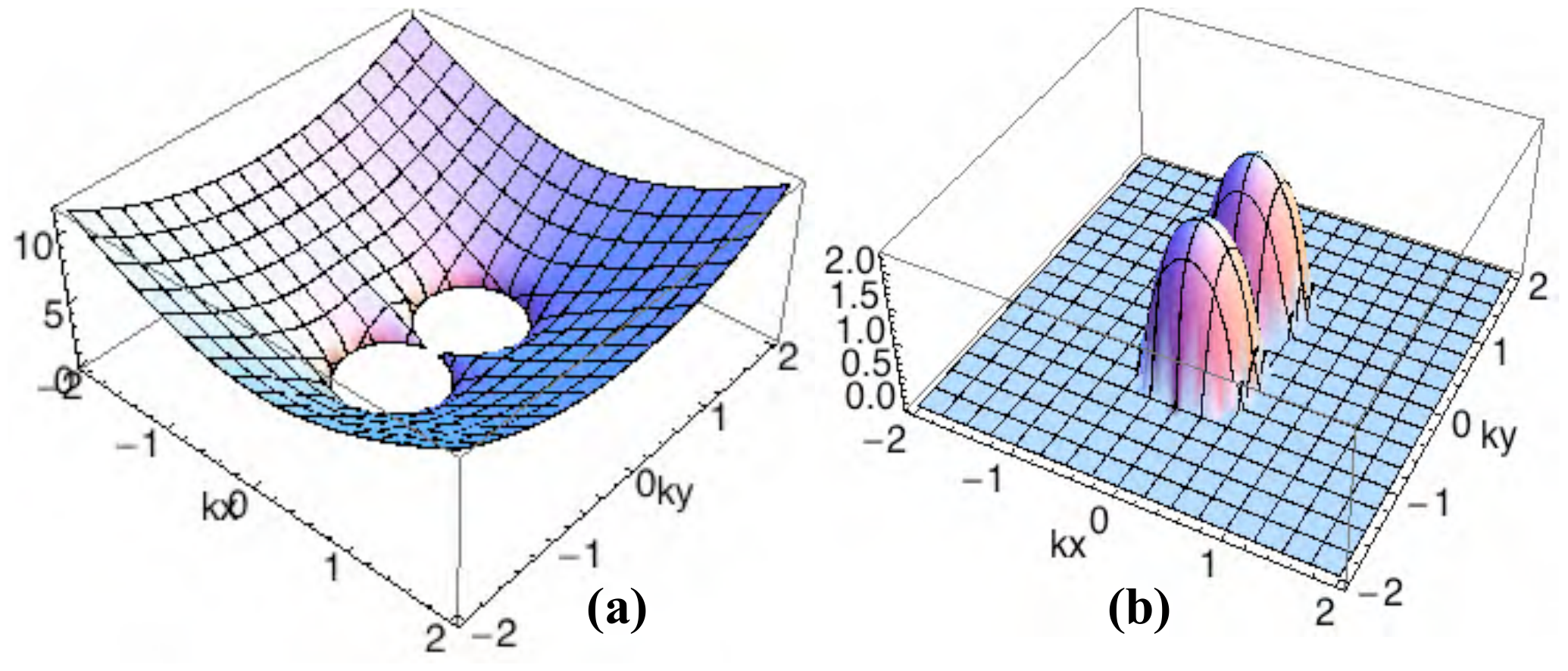}
\caption{(Color online) An example of real (a) and imaginary (b) part of unstable branch of excitation spectrum with SO coupling.  \label{BG}}
\end{figure}

Suppose $H_{{\bf k}}$ can be diagonalized by a $\Omega_{{\bf k}}$, and the diagonal values of $\Omega^\dag_{{\bf k}}H_{{\bf k}}\Omega_{{\bf k}}$ are denoted by $\lambda_i$. $\lambda_i$ satisfies the equation $\text{Det}[H_{{\bf k}}-\lambda_i A]=0$, which gives two solutions denoted by $\lambda_{\uparrow{\bf k}}$ and $\lambda_{\downarrow{\bf k}}$. The Bogoliubov Hamiltonian becomes $\hat{H}_{\text{Bg}}=\sum_{{\bf k},\sigma}\lambda_{\sigma{\bf k}}\beta^\dag_{\sigma{\bf k}}\beta_{\sigma{\bf k}}$. We find that one excitation branch always has a positive imaginary part in a large regime in momentum space, {\it even for infinitesimal small SO coupling}. Therefore, suppose a BEC is initially prepared in the ${\bf k}=0$ state, as SO coupling is turned on, some modes will exponentially grow. Therefore, a conventional BEC on ${\bf k}=0$ mode is unstable with SO coupling, which arises the question that what is the actual ground state of a spinor BEC with SO coupling.

In addition, we note that the spectrum shown in Fig. \ref{BG} breaks spatial rotation symmetry. This is also an effect of SO coupling. Since the Hamiltonian of Eq. \ref{single} only possess a symmetry of simultaneous rotation of both spin and space, namely $k_x+ik_y\rightarrow e^{i\theta}(k_x+ik_y)$ and $\sigma_x\pm i\sigma_y\rightarrow e^{\pm i\theta}(\sigma_x\pm i\sigma_y)$. However, the Bose condensation locks the relative phase between different spin components and therefore breaks spin rotation symmetry. Consequently, this symmetry breaking manifests itself in real space.

\begin{figure}[bp]
\includegraphics[height=1.8in, width=1.8in]
{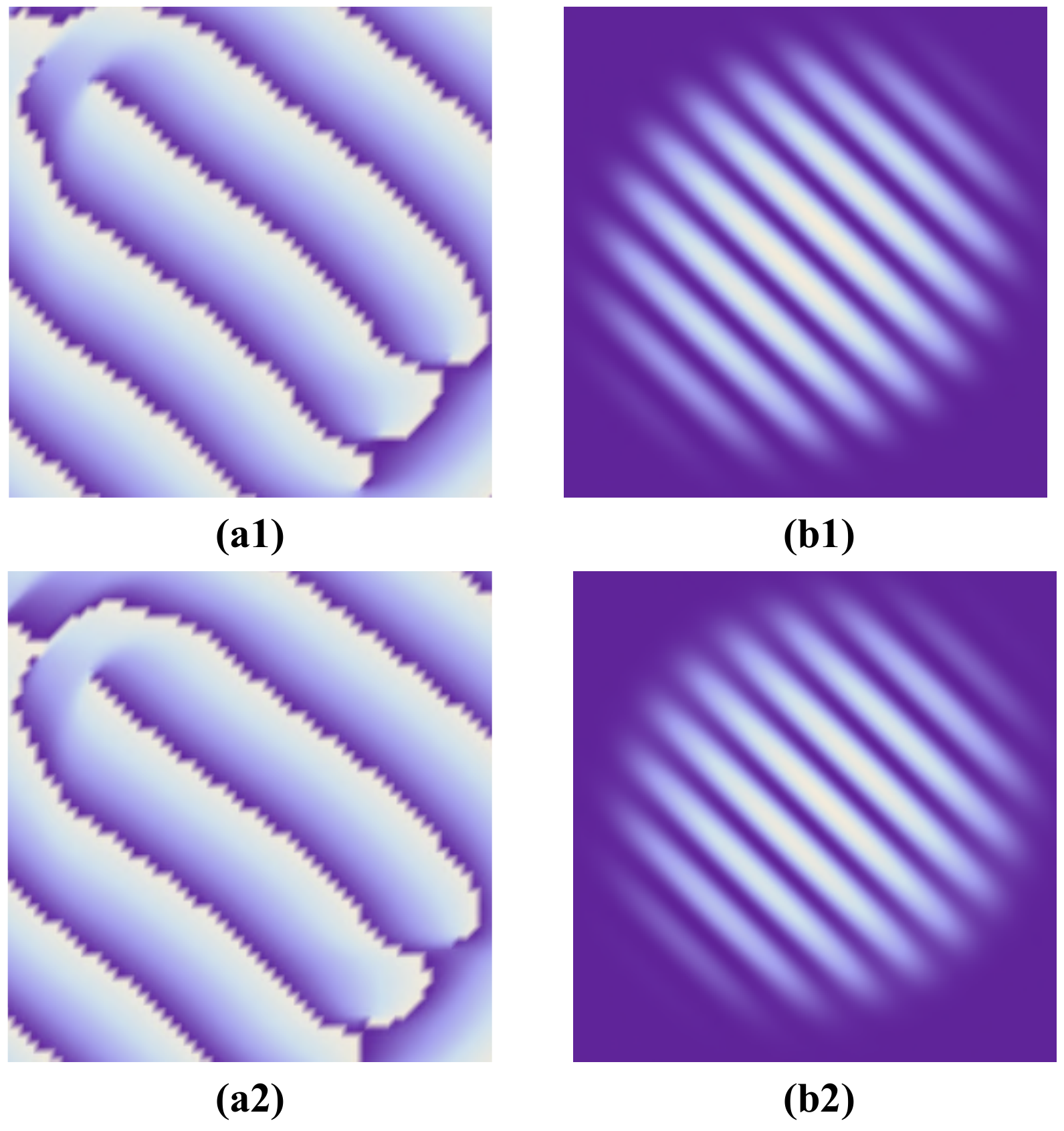} \caption{(Color online) Numerical results for spin-$1/2$ case. a1 and a2 show the phase of condensate wave function of both spin-up (a1) and down (a2) component increase from $-\pi$ (grey regime) to $\pi$ (dark regime) periodically in the PW regime; b1 and b2 show the density of both spin-up (b1) and down (b2) component oscillate periodically in the SW regime. $\gamma=c_2/c_0=0.4$ for a1 and a2, and $\gamma=0.1$ for b1 and b2.  \label{spinhalf}}
\end{figure}

{\it Numerical Simulation for Spin-$1/2$ Case:} We implement the mean-field approximation and numerically look for the condensate wave function that can minimize Gross-Pitaevskii energy using imaginary time evolution method. For spin-$1/2$ case the Gross-Pitaevskii energy is written as
\begin{eqnarray}
\mathcal{E}=&&\int d^2{\bf r}\left\{\sum\limits_{\sigma=\uparrow,\downarrow}\varphi^*_{\sigma}\left(-\frac{\hbar^2}{2m}\nabla^2+\frac{1}{2}m\omega^2 r^2\right)\varphi_{\sigma}\right.\nonumber\\
&&\left.+\frac{\hbar\kappa}{m} \left[\varphi^*_{\uparrow}(-i \partial_x- \partial_y)\varphi_{\downarrow}+\varphi^*_{\downarrow}(-i \partial_x+\partial_y)\varphi_{\uparrow}\right]\right.\nonumber\\
&&\left.+\frac{c_0}{2}\left(|\varphi_\uparrow|^2+|\varphi_\downarrow|^2\right)^2+\frac{c_2}{2}\left(|\varphi_\uparrow|^2-|\varphi_\downarrow|^2\right)^2\right\} \label{Eng}
\end{eqnarray}
What we find are shown in Fig. \ref{spinhalf}. For $\gamma=c_2/c_0>0$, the densities of both component have no particular structure, while the phase of both components behave as a plane wave, as shown in Fig. \ref{spinhalf}(a) and named as ``PW" phase. Time reversal symmetry is broken in this phase. For $\gamma<0$, the wave functions of both components behave as standing waves and lead to periodic density modulation of both components, as shown in Fig. \ref{spinhalf}(b) and named as ``SW" phase. Time reversal symmetry is preserved. Moreover, the higher density regime of spin-up component coincides with the lower density regime of spin-down component, which represents a microscopic phase separation, and also represents a spin stripe state. 

Though in the numerical simulation, we have included a very weak harmonic trap which helps to avoid artifact from a sharp boundary and also simulates the practical situation in cold atom experiment, the results can be understood from a homogeneous case. With SO coupling, the single-particle spectrum becomes $E_{\pm{\bf k}}=(\hbar^2 {\bf k}^2\pm 2\kappa\hbar |{\bf k}|)/(2m)$, where $\pm$ denotes different helicity (spin parallel or anti-parallel to wave vector). The single particle ground state is in the negative helicity branch with $|{\bf k}|=\kappa/\hbar$, and the wave function is given by
\begin{equation}
\phi_{{\bf k}}=\frac{1}{\sqrt{2}}e^{i{\bf k}{\bf r}}\left(\begin{array}{c}1 \\-e^{i\varphi_{{\bf k}}}\end{array}\right).
\end{equation}
where $\varphi_{{\bf k}}=\text{arg}(k_x+ik_y)$ and $\varphi_{{\bf -k}}=\varphi_{{\bf k}}+\pi$.
Let us first consider a simple case that assumes the condensate wave function is a superposition of two opposite wave vector state as
\begin{eqnarray}
\varphi=\left(\begin{array}{c}\varphi_{\uparrow} \\ \varphi_{\downarrow}\end{array}\right)= \frac{\alpha_1}{\sqrt{2}}e^{i{\bf k}{\bf r}}\left(\begin{array}{c}1 \\-e^{i\varphi_{{\bf k}}}\end{array}\right)
+\frac{\alpha_2}{\sqrt{2}}e^{-i{\bf k}{\bf r}}\left(\begin{array}{c}1 \\-e^{i\varphi_{{\bf -k}}}\end{array}\right)\nonumber
\end{eqnarray}
Using this ansatz to minimize the interaction energy, it is easy to find, for $c_2<0$, it favors $\alpha_1=\alpha_2=1/\sqrt{2}$, therefore, $\varphi_{\uparrow}\sim \cos{\bf k}{\bf r}$ and $\varphi_{\downarrow}\sim i\sin {\bf k}{\bf r}$. While for $c_2>0$, it favors the case $\alpha_1=1$, $\alpha_2=0$ or $\alpha_1=0$, $\alpha_2=1$, namely, the wave function is a single plane wave. 

The next question is whether there will be more than one single ${\bf k}$ state, or a pair of $\{{\bf k},{\bf -k}\}$ states entering the condensate wave function $\varphi$. In general, one shall assume a superposition of all states in the degenerate circle
\begin{equation}
\varphi=\int d\varphi_{{\bf k}}\frac{\alpha_{{\bf k}}}{\sqrt{2}}e^{i{\bf k}{\bf r}}\left(\begin{array}{c}1 \\-e^{i\varphi_{{\bf k}}}\end{array}\right). \label{ansatz}
\end{equation}
where the amplitude of ${\bf k}$ is fixed at $\kappa/\hbar$ to minimize the single particle energy. For instance, if $\alpha_{{\bf k}}$ is independent of the angle of ${\bf k}$, one can obtain
\begin{equation}
\varphi=\frac{1}{\sqrt{2}}\left(\begin{array}{c}\pi J_0(|{\bf k}||{\bf r}|) \\ i\pi J_1(|{\bf k}|{\bf r}|)e^{i\theta}\end{array}\right).
\end{equation}
where $\theta$ is the angle of ${\bf r}$. This is a symmetric skyrmion solution, which has also been proposed by Ref. \cite{Wu}. However, if one substitutes the ansatz Eq. \ref{ansatz} into the energy function Eq. \ref{Eng} and minimize energy with respect to all $\alpha_{{\bf k}}$, we can find the most favorable solution is always that $\alpha_{{\bf k}}$ is non-zero either for a single ${\bf k}$ or for a pair of $\{{\bf k},{\bf -k}\}$, and we do not find a parameter regime in which the condensate wave function contains more than two wave vector components.

{\it Numerical Simulation for Spin-$1$ Case:} We now move to study spin-$1$ case, whose energy functional is given by
\begin{eqnarray}
\mathcal{E}=&&\int d^3{\bf r}\left\{\sum\limits_{\sigma=1,0,-1}\varphi^*_{\sigma}\left(-\frac{\hbar^2}{2m}\nabla^2+\frac{1}{2}m\omega^2 r^2\right)\varphi_{\sigma}\right.\nonumber\\
&&\left.+\frac{\hbar\kappa}{m} \left[\varphi^*_{1}(-i \partial_x- \partial_y)\varphi_{0}+\varphi^*_{0}(-i \partial_x-\partial_y)\varphi_{-1}+\text h.c. \right]\right.\nonumber\\
&&\left.+\frac{c_0}{2}\left(|\varphi_1|^2+|\varphi_0|^2+|\varphi_{-1}|^2\right)^2\right.\nonumber\\
&&\left.+\frac{c_2}{2}\left[\left(|\varphi_1|^2-|\varphi_{-1}|^2\right)^2+2\left|\varphi_1^*\varphi_0+\varphi^*_0\varphi_{-1}\right|^2\right]\right\}
\end{eqnarray}
The results from numerical simulation are displayed in Fig. \ref{spin1}. As shown in Fig. \ref{spin1}, for $\gamma>0$ it is ``SW" phase and for $\gamma<0$ it is ``PW" phase.

The result of spin-$1$ can be understood with similar analysis above. The wave function for single particle ground state is now
\begin{equation}
\phi_{{\bf k}}=\frac{1}{2}\left(\begin{array}{c} 1 \\ -\sqrt{2}e^{i\varphi_{{\bf k}}} \\ e^{i2\varphi_{{\bf k}}} \end{array}\right)e^{i{\bf k}{\bf r}}
\end{equation}
We shall also consider a superposition as $\alpha_1 \phi_{{\bf k}}+\alpha_2 \phi_{{\bf -k}}$. $c_0$ term is independent of $\alpha_{1,2}$, however, for $\alpha_1=\alpha_2=1/\sqrt{2}$, $\langle {\bf S}\rangle^2=0$ and it is favored when $c_2>0$. This state is also a special case of so-called ``polar" or ``nematic" phase in the discussion of spin-$1$ BEC \cite{Ho}; while for $\alpha_1=0$, $\alpha_2=1$ or $\alpha_1=1$, $\alpha_2=0$, $\langle {\bf S}\rangle^2=1$ and it is favored when $c_2<0$. This is also called ``ferromagnetic" phase.

{\it Domains and domain wall:} Our numerical simulation also finds long-lived metastable states with domains. For instance, if we start with a random initial configuration, most cases the imaginary time evolution leads to a state in which the system often splits into two domains. In the ``PW" phase, the system is locally a single plane wave state in each domain, and the wave vector is opposite between two domains. This is very similar to the situation of ferromagnetism, where one always finds a ferromagnetic state made up with locally magnetized domains. 

\begin{figure}[tbp]
\includegraphics[height=1.7in, width=2.3in]
{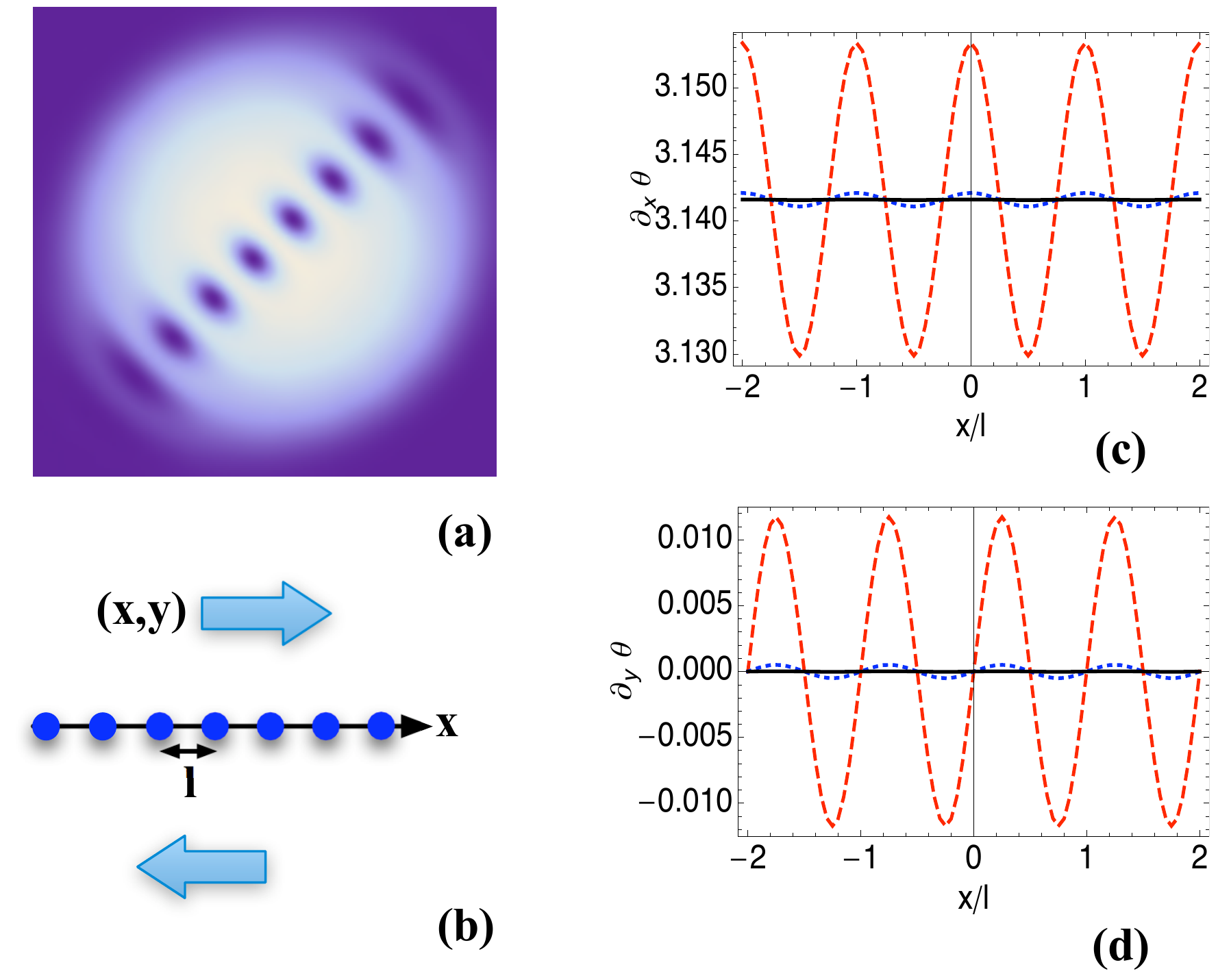} \caption{(Color online) (a) Density of up (or down) component with two domains in the ``PW" phase. (b) A schematic of vortex array as a domain wall between two regimes with opposite wave vector. (c-d) Consider array of vortices separated by distance $l$ located in $x$-axes, $\partial_x\theta$ (c) and $\partial_y\theta$ (d) (in unit of $1/l$) at position $(x,y)$ as a function of $x/l$ is shown for $|y|=l$ (red dashed line), $|y|=1.5l$ (blue dotted line) and $|y|=2l$ (black solid line). $\theta$ is the phase of condensate wave function. \label{domain}}
\end{figure}

In presence of domains, we find that both up and down components of the condensate wave function contain an array of vortices, as shown in Fig. \ref{domain}(a). We have checked that the vorticity of all vortices are the same, and the vortices in different components locate alternately. The vortex array plays the role as a domain wall.  Consider an array of vortices with same vorticity, located at $x=nl$ and $y=0$, where $n$ are integers, in a uniform superfluid the gradient of superfluid phase $\vec{\partial}\theta$  at $(x,y)$ is given by
\begin{eqnarray}
\partial_x \theta&=&\frac{1}{l}\sum\limits_{n=-\infty}^{+\infty}\frac{y/l}{(x/l+n)^2+(y/l)^2}; \nonumber\\
\partial_y\theta&=&-\frac{1}{l}\sum\limits_{n=-\infty}^{+\infty}\frac{x/l+n}{(x/l+n)^2+(y/l)^2}.
\end{eqnarray}
$\vec{\partial}\theta$ as a function of $(x,y)$ is shown in Fig. \ref{domain} (c-d). As one can see, when $|y|>l$, $\partial_x\theta\rightarrow \pi/l$ is a constant and $\partial_y\theta\rightarrow 0$. Hence the system is locally a plane wave state. To minimize the single particle energy, one requires $l=\pi\hbar/\kappa$, namely, the vortex line density increases as the increase of SO coupling. 

Hence, we have established the conclusion \textbf {(I-IV)} summarized above within a mean-field theory. In future studies we will include quantum fluctuations. Due to the degeneracy of single particle ground states, quantum fluctuation, in particular, the fluctuation of rotation mode, may lead to fragmentation. However, it is known that fragmented state is usually very fragile for a realistic system with large number of bosons and is not stable against external perturbations. In this case, effects such as anisotropy of trapping potential will break the spatial rotational symmetry and pin the direction of plane wave or density stripe, and prevent the fluctuation of rotation mode from restoring the symmetry. Mean-field results become more stable and the predictions of this work can be verified experimentally very soon.

{\it Acknowledgment}: HZ is supported by the Basic Research Young
Scholars Program of Tsinghua University, NSFC Grant No. 10944002.

\end{document}